\documentclass[12pt, a4paper]{article}
\usepackage{graphicx}
\usepackage{amssymb,amsmath,bm,mathrsfs}
\usepackage{xcolor}
\usepackage{float}
\usepackage{physics}
\usepackage[colorlinks=true ,urlcolor=blue ,anchorcolor=blue ,citecolor=blue ,filecolor=blue ,linkcolor=blue ,menucolor=blue ,pagecolor=blue ,linktocpage=true ,pdfproducer=medialab ,pdfa=true]{hyperref}
\usepackage[normalem]{ulem}
\usepackage[top=30truemm,bottom=30truemm,left=25truemm,right=25truemm]{geometry}

\usepackage{comment}

\begin{document}
\begin{titlepage}

\vskip 2cm
\begin{center}

{\Large 
{\bf
Revisiting Bino-Slepton Coannihilation Dark Matter in Light of Recent Experimental Results
}
}

\vskip 1.5cm

Koichi~Hamaguchi$^{a,b}$\footnote{
\href{mailto:hama@hep-th.phys.s.u-tokyo.ac.jp}{\tt
 hama@hep-th.phys.s.u-tokyo.ac.jp}},
Atsuya Niki$^a$\footnote{
\href{mailto:niki@hep-th.phys.s.u-tokyo.ac.jp}{\tt niki@hep-th.phys.s.u-tokyo.ac.jp}}, and
Kwok Hei To$^a$\footnote{
\href{mailto:alethea.t.0124@hep-th.phys.s.u-tokyo.ac.jp}{\tt alethea.t.0124@hep-th.phys.s.u-tokyo.ac.jp}}

\vskip 0.8cm

{\it $^a$Department of Physics, University of Tokyo, Bunkyo-ku, Tokyo
 113--0033, Japan} \\[2pt]
 {\it $^b$Kavli IPMU (WPI), University of Tokyo, Kashiwa, Chiba
  277--8583, Japan} \\[2pt]

\date{\today}

\vskip 1.5cm

\begin{abstract}

Despite being a simple and well-motivated thermal relic scenario, coannihilation dark matter (DM) has remained largely unexplored experimentally due to the difficulty of probing its nearly degenerate mass spectrum.
Recent LHC searches, however, have significantly improved the sensitivity to such compressed spectra, motivating a reassessment of the viable parameter space.
We revisit the bino-slepton coannihilation scenario in supersymmetric (SUSY) models, incorporating the latest experimental results.
We first focus on the minimal scenario, in which only the bino-like neutralino and left- or right-handed sleptons are light (${\cal O}(100)$ GeV), with all other SUSY particles decoupled. 
We find that the dark matter mass is constrained to be in the range of about 170--420 GeV (130--430 GeV) for left-handed (right-handed) slepton coannihilation, with lower bounds set by recent LHC searches.
We then investigate scenarios with light higgsino, where direct detection experiments impose strong constraints on the higgsino mass.
We also discuss the implications of these constraints for the muon $g-2$ in the so-called BHR, BHL, and BLR scenarios with coannihilation DM, and find that the combined LHC and LZ limits constrain the SUSY contribution to $|\Delta a_\mu^{\rm SUSY}|\lesssim 1.2\times10^{-9}$.
\end{abstract}

\end{center}
\end{titlepage}

\section{Introduction}

Astrophysical and cosmological observations have provided strong support for the existence of dark matter (DM), which accounts for a significant portion of the Universe's mass density, measured as $\Omega_{\rm DM}h^2=0.120$~\cite{Planck:2018vyg}.
However, the Standard Model (SM) of particle physics, despite its extraordinary success in describing known particles and forces, does not provide a viable candidate for DM. 

Among various frameworks proposed for explaining dark matter, the thermal relic scenario stands out as particularly appealing. 
It naturally accounts for the observed dark matter abundance through the freeze-out mechanism, a robust and well-motivated process in the early Universe. 
Within this framework, coannihilation~\cite{Griest:1990kh} is one of the simplest viable mechanisms, effectively reducing the dark matter relic abundance through interactions involving nearly degenerate partner particles. 
Although the relevant parameter region has long remained difficult to probe experimentally because of the near-degenerate mass spectrum, recent advances in LHC searches have significantly improved the sensitivity to such compressed spectra, motivating a reassessment of the viable parameter space.

In this paper, we revisit the scenario of bino-slepton coannihilation DM in supersymmetric (SUSY) models. 
SUSY is an attractive extension of the Standard Model, providing a stable DM candidate as the lightest SUSY particle (LSP) in the minimal supersymmetric Standard Model (MSSM) with R-parity~\cite{Martin:1997ns}.
In the bino-slepton coannihilation scenario, the nearly degenerate bino and slepton masses allow coannihilation to reduce the thermal relic abundance to the observed value.

We first consider the minimal slepton coannihilation scenarios where only the pure bino and the left-handed (or right-handed) sleptons are light, assuming all other SUSY particles and heavy Higgs bosons are decoupled. 
We show that the recent LHC results on the search for SUSY particles with compressed mass spectra~\cite{ATLAS:2025evx,CMS:2025ttk} place stringent constraints on such models, restricting the dark matter mass to a narrow range, approximately 170--420 GeV and 130--430 GeV for left-handed and right-handed slepton coannihilation, respectively, with the lower bounds set by the latest LHC results.

Next, we investigate a next-to-minimal scenario where, in addition to the bino and slepton masses, the higgsino parameter is also small. Such a small higgsino parameter is motivated by the naturalness of the electroweak symmetry breaking~\cite{Barbieri:1987fn,Kitano:2006gv,Baer:2012up}.
In this case, the recent results from the direct detection experiment LZ~\cite{LZ:2024zvo} impose stringent constraints on the higgsino
mass.

We also discuss the implications of these constraints
for the muon anomalous magnetic moment ($g-2$) in the so-called BHR, BHL,
and BLR scenarios. We show that, in these scenarios, 
the LHC and dark matter direct-detection constraints restrict the
SUSY contribution to the muon $g-2$ to be $|\Delta a_\mu^{\rm SUSY}| \lesssim 1.2\times 10^{-9}$.

This paper is organized as follows. Section~\ref{sec:Bino-Slepton} considers the minimal slepton coannihilation models and discusses the relic-abundance requirement and the current LHC constraints.
Section~\ref{sec:Bino-Slepton-Higgsino} explores the slepton coannihilation model with a light higgsino and its phenomenological implications, including dark matter direct detection and muon $g-2$.
Finally, Section~\ref{sec:summary} offers a summary and discussion of this paper.

\section{Minimal slepton coannihilation model}
\label{sec:Bino-Slepton}

\subsection{Model setup}
\label{subsec:Bino-Slepton-model}

In this section, we focus on the minimal slepton coannihilation scenarios, in which the bino-like neutralino is the LSP and the left- or right-handed sleptons are slightly heavier than the bino.
We refer to these as the BL and BR models, respectively.
All other SUSY particles and heavy Higgs bosons are assumed to be sufficiently heavy so that they do not affect the dark matter phenomenology.

For each of the BL and BR models, we consider three flavor patterns, depending on which slepton species is light:
(i) only the selectron, (ii) only the smuon, and (iii) all three generations being degenerate.
In total, we analyze six scenarios.

The relevant parameters for the present analysis are
\begin{align}
M_1,\ M_{L_i/R_i},\ \tan\beta,\ \mu,
\label{eq:MSSMparameters}
\end{align}
where $M_1$ is the bino soft mass, $M_{L_i}$ and $M_{R_i}$ ($i=1,2,3$) are the soft masses of the left- and right-handed sleptons of each generation, $\mu$ is the higgsino mass parameter, and $\tan\beta$ is the ratio of the vacuum expectation values of the up- and down-type Higgs~\cite{Martin:1997ns}.
In this section, we fix $\mu = 5~\mathrm{TeV}$ and vary $M_1$ and $M_{L_i/R_i}$ according to the flavor patterns discussed above. For the BL models:
\begin{itemize}
  \item[(i)] Only the selectron is light: $M_1$ and $M_{L_1}$ are varied, while $M_{L_{2,3}} = M_{R_i}= 5~\mathrm{TeV}$;
  \item[(ii)] Only the smuon is light: $M_1$ and  $M_{L_2}$ are varied, while $M_{L_{1,3}}=M_{R_i} = 5~\mathrm{TeV}$;
  \item[(iii)] All three generations are degenerate: $M_1$ and $M_{L_1}=M_{L_2}=M_{L_3}\equiv M_L$ are varied, while $M_{R_i}= 5~\mathrm{TeV}$.
\end{itemize}
The parameters of the BR models are varied in an analogous manner, with the roles of left- and right-handed sleptons interchanged.
The other SUSY particles and the heavy Higgs bosons are decoupled by setting their soft masses to 5~TeV, and we neglect the trilinear $A$-terms and the complex phases of the soft parameters.

The lightest neutralino mass is well approximated by $m_{\tilde{\chi}^0_1}\simeq M_1$, 
since the higgsino and wino are decoupled.
The slepton masses are mainly determined by their soft masses, $M_{L_i/R_i}$, with minor corrections that depend on $\tan\beta$ and $\mu$.
The physical masses of the charged sleptons are the eigenvalues of the mass matrix~\cite{Martin:1997ns}
\begin{align}
    \begin{pmatrix}
    M_{L_i}^2 + m_Z^2\left(-\frac{1}{2}+ \sin^2\theta_W\right)\cos2\beta & -m_{\ell_i}\mu\tan\beta \\ -m_{\ell_i}\mu\tan\beta & M_{R_i}^2 - m_Z^2\sin^2\theta_W\cos2\beta
    \end{pmatrix}.
    \label{eq:slepton_mass_matrix}
\end{align}
The terms proportional to $m_Z^2$ arise from the $D$-term contributions, and the off-diagonal terms are from the $F$-term contributions.
Here, $\theta_W$, $m_Z$, and $m_{\ell_i}$ ($\ell_i = e, \mu, \tau$) are the weak mixing angle, the $Z$-boson mass, and the lepton masses, respectively. We have neglected $m_{\ell_i}^2$ terms in the diagonal components, for simplicity.
For the first and second generations, the off-diagonal elements are negligible, so that the mass eigenstates are well approximated by the gauge eigenstates.
In this case, we denote the corresponding physical slepton masses by
$m_{\tilde{\ell}_L}$ and $m_{\tilde{\ell}_R}$ ($\ell=e,\mu$).
For the third generation, the off-diagonal elements can provide sizable corrections depending on the value of $\mu \tan\beta$.
The sneutrino masses are given by $m_{\tilde{\nu_i}}^2 = M_{L_i}^2 + \frac{1}{2}m_Z^2\cos2\beta$.

\subsection{Relic Abundance}
\label{subsec:Bino-Slepton-omega}

In coannihilation scenarios, the thermal relic abundance of DM is determined mainly by the masses of the DM and its coannihilating partner.
Their small mass difference, typically $\sim m_{\rm DM}/20$, enhances the coannihilation process in the early Universe, resulting in a thermal relic density consistent with the observed value.
In the BL model, the bino DM mainly coannihilates with the sneutrinos, which are lighter than the charged sleptons by $m_{\tilde{\ell}_L}^2-m_{\tilde{\nu}}^2\simeq m_W^2|\cos2\beta|$ when $\tan\beta>1$. Here, $m_W$ is the $W$-boson mass. In the BR model, the DM coannihilates with the charged sleptons.

\begin{figure}[t]
  \centering
  \includegraphics[width=7cm,angle=0]{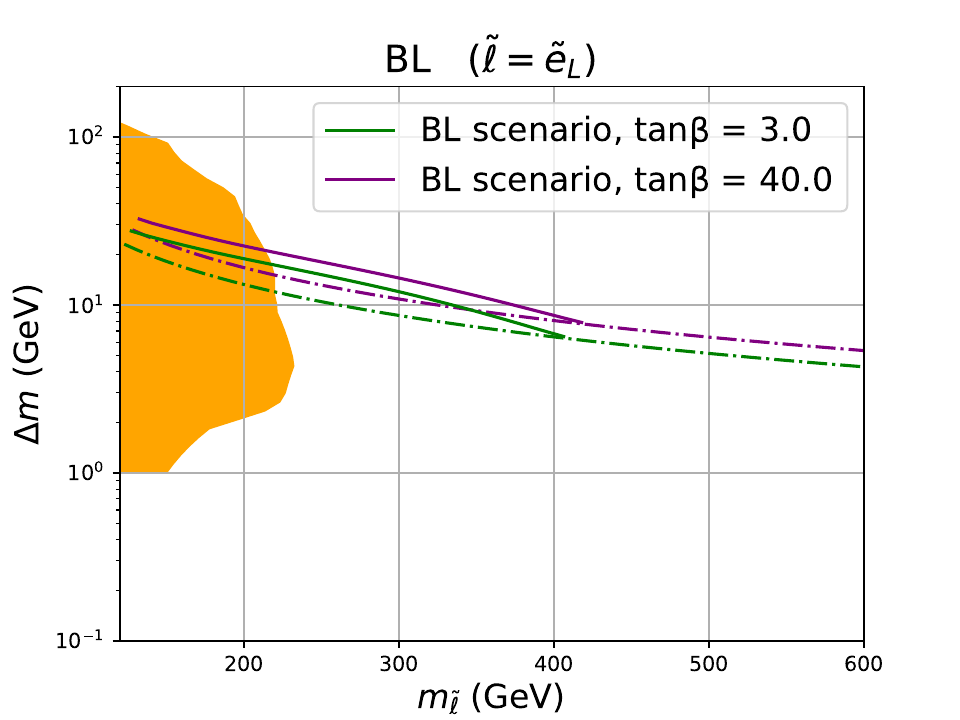}
  \includegraphics[width=7cm,angle=0]{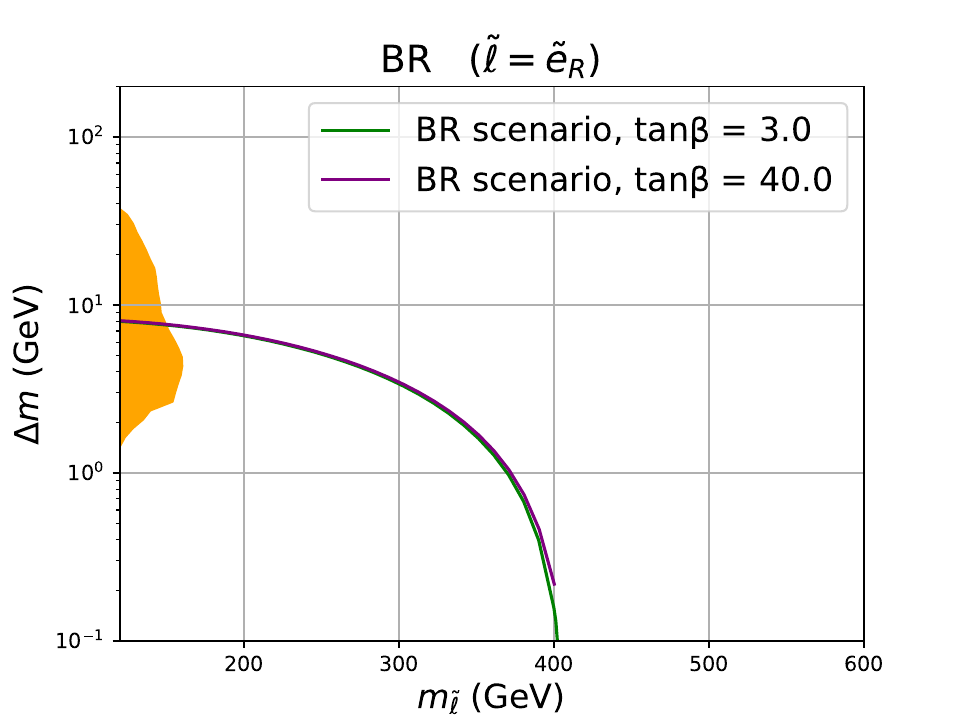}
  \\
  \includegraphics[width=7cm,angle=0]{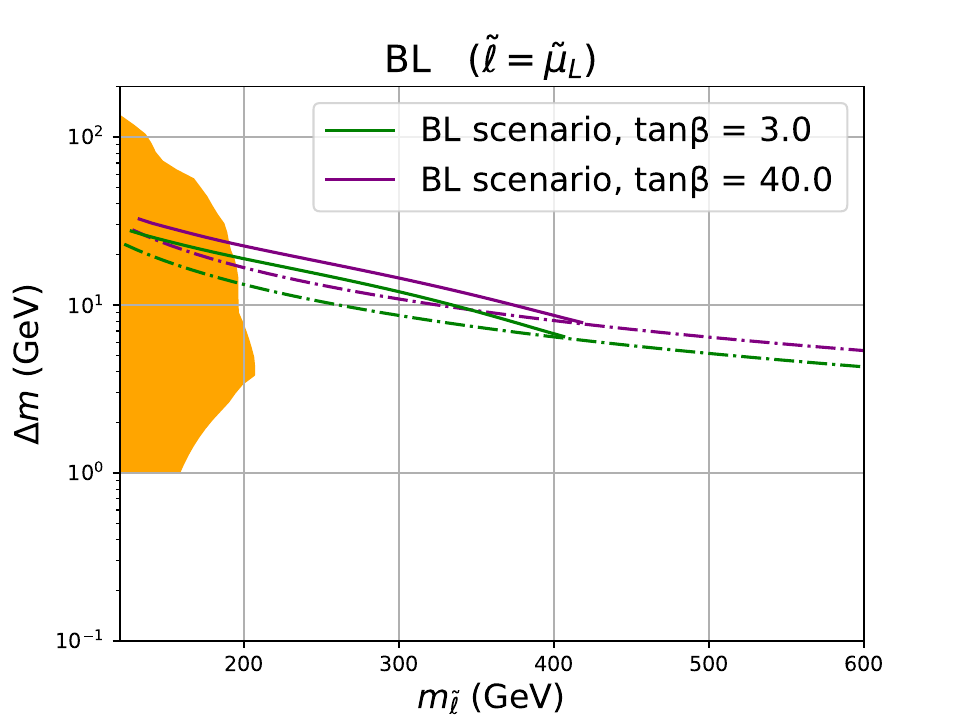}
  \includegraphics[width=7cm,angle=0]{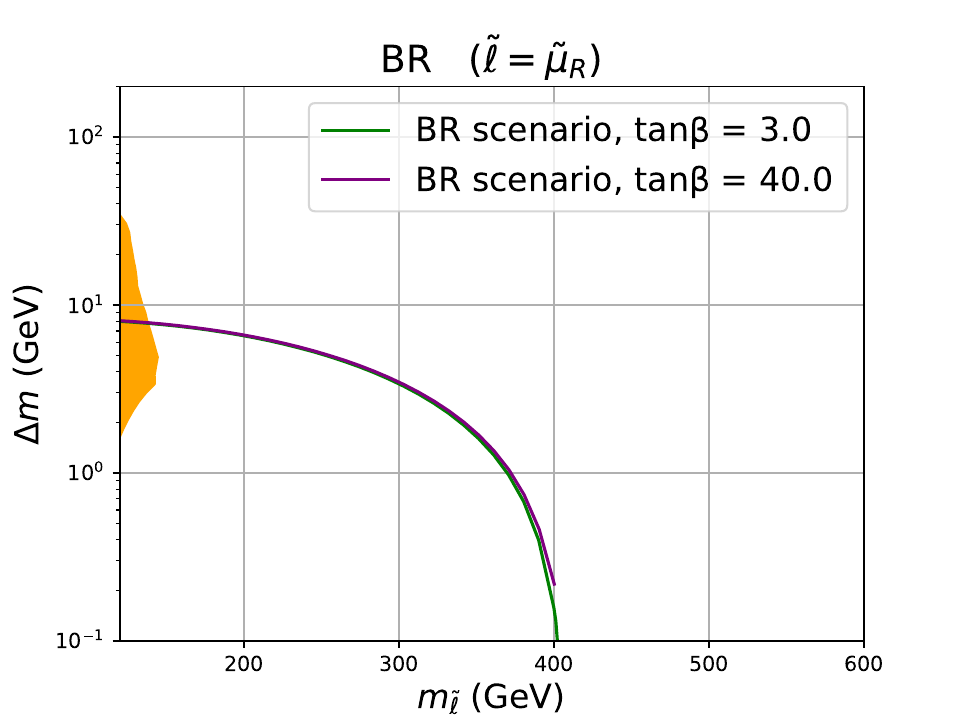}
  \caption{
  The relic abundance contours and the LHC constraints in the $(m_{\tilde{\ell}}, \Delta m)$ plane in the minimal slepton coannihilation scenario for the BL (left) and BR (right) models.
  The upper panels correspond to the cases where the bino coannihilates with the selectron, and the lower panels to those with the smuon.
  Here, $\Delta m = m_{\tilde{\ell}}-m_{\tilde{\chi}^0_1}$ and $m_{\tilde{\ell}}$ denotes the lighter selectron or smuon mass.
The solid lines show the contours of $\Omega_{\tilde{\chi}^0_1}h^2 = \Omega_{\rm DM}h^2 = 0.120$.
The orange-shaded regions represent the exclusion limits from the CMS soft-lepton search~\cite{CMS:2025ttk}.
In the BL scenario (left panels), the dot-dashed line indicates the boundary where the LSP changes from the neutralino to the slepton; the region below the line corresponds to a slepton LSP.
The green and purple lines correspond to $\tan\beta = 3$ and $40$, respectively.
  }
  \label{fig:BR_BL}
\end{figure}

\begin{figure}[t]
  \centering
  \includegraphics[width=7cm,angle=0]{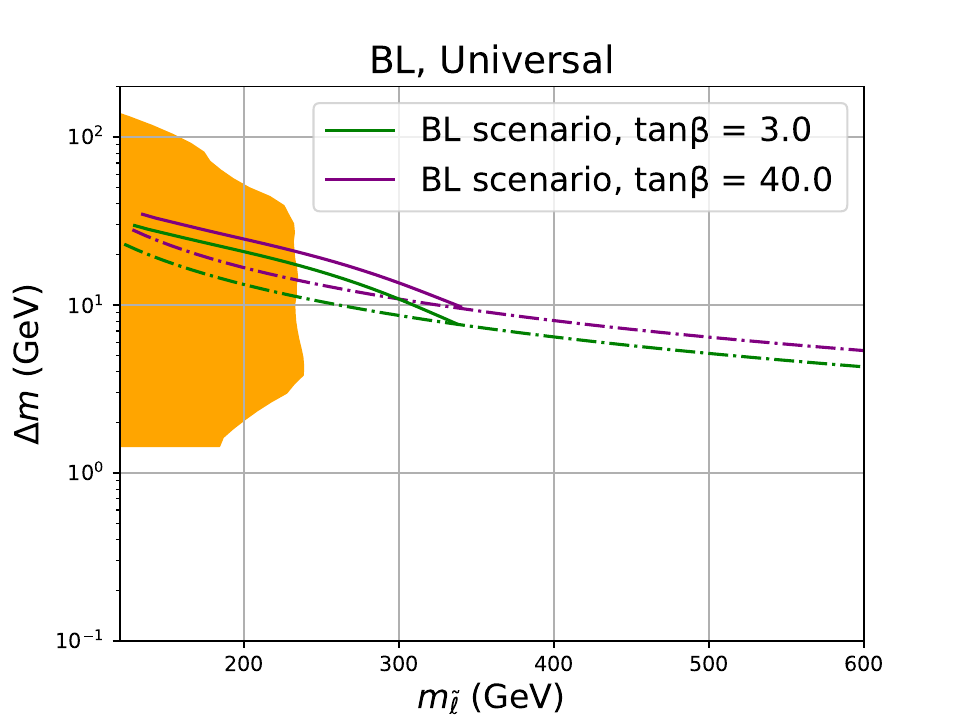}
  \includegraphics[width=7cm,angle=0]{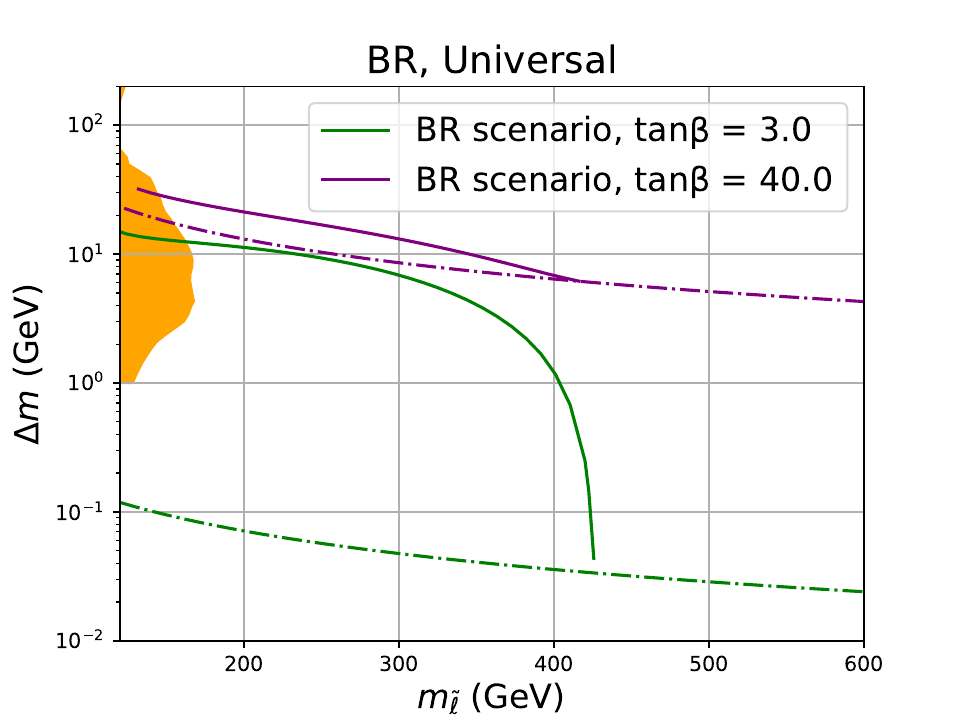}
  \caption{The same as Fig.~\ref{fig:BR_BL}, but for the flavor-universal slepton masses ($M_{L_1}=M_{L_2}=M_{L_3}$ and $M_{R_1}=M_{R_2}=M_{R_3}$).
  The solid and dot-dashed lines denote the relic-abundance contours and the LSP boundaries, respectively, while the orange-shaded regions indicate the LHC exclusions.
  Note that the x-axis shows the lighter selectron/smuon mass.
  }
  \label{fig:BR_BL_Universal}
\end{figure}

We calculate the DM relic abundance, $\Omega_{\tilde{\chi}^0_1}$, using the public package ${\tt micrOMEGAs 6.2.3}$~\cite{Belanger:2001fz,Alguero:2023zol}.
In Figs.~\ref{fig:BR_BL} and \ref{fig:BR_BL_Universal}, we plot the contours showing $\Omega_{\tilde{\chi}^0_1}h^2=\Omega_{\rm obs}h^2=0.120$ in the $(m_{\tilde{\ell}}, \Delta m)$-plane for each model described in Sec.\ref{subsec:Bino-Slepton-model}.
Hereafter, $\tilde{\ell}$ denotes the lighter selectron or smuon, and
\begin{align}
\Delta m = \Delta m(\tilde{\ell},\tilde{\chi}^0_1) = m_{\tilde{\ell}}-m_{\tilde{\chi}^0_1}.
\end{align}
We choose $\Delta m(\tilde{\ell},\tilde{\chi}^0_1)$ as the variable since it allows for a clearer comparison with the LHC constraints, as discussed in the next subsection.

In Fig.~\ref{fig:BR_BL}, the upper and lower panels represent the selectron and smuon coannihilation cases, respectively, while Fig.~\ref{fig:BR_BL_Universal} shows the flavor-universal case. The left and right panels correspond to the BL and BR models, respectively.
The solid lines indicate the contours of $\Omega_{\tilde{\chi}^0_1}h^2 = 0.120$, and the dot-dashed lines mark the LSP boundaries; the region below the line corresponds to a slepton LSP.
The orange-shaded regions show the exclusion limits from the LHC soft-lepton search and will be discussed in the next subsection.

In the BL scenario, the bino DM mainly coannihilates with sneutrinos, which are lighter than the charged sleptons by approximately $m_W^2|\cos 2\beta|$.
As a result, the mass splitting between the DM and the charged slepton that yields the observed relic abundance is somewhat larger than in the BR scenario, where the bino coannihilates with charged sleptons.
The small difference between the selectron/smuon cases and the flavor-universal case in the BL model, 
as seen in Figs.~\ref{fig:BR_BL} and \ref{fig:BR_BL_Universal}, arises from the left–right mixing in the stau mass matrix.
For the selectron and smuon BR cases, the slepton masses are practically independent of $\tan\beta$.
In contrast, a $\tan\beta$ dependence and the appearance of the LSP boundary are observed in the flavor-universal BR case, due to the left–right mixing in the stau mass matrix.

Note that, as shown in Figs.~\ref{fig:BR_BL} and \ref{fig:BR_BL_Universal}, the charged slepton masses consistent with the observed relic abundance are bounded from above, $m_{\tilde{\ell}}\lesssim (340$–$430)$ GeV.
This upper bound arises because, at larger DM masses, even a vanishing mass splitting ($\Delta m \simeq 0$) would lead to an overabundant relic density.

\subsection{LHC constraints}
\label{subsec:Bino-Slepton-exp}

Experimental probes of coannihilation DM scenarios have long been challenging, especially at the LHC.
The compressed mass spectrum between the bino-like DM and the slepton leads to soft decay products, making it difficult to distinguish the signal from the Standard Model backgrounds.
Until recently, previous LHC analyses (e.g., \cite{ATLAS:2019lff,ATLAS:2019gti,ATLAS:2022hbt,ATLAS:2014zve}) had not set limits on the parameter space consistent with the observed thermal relic abundance, beyond the LEP exclusion region of $m_{\tilde{\ell}}<100$ GeV~\cite{LEP:2004hbt}.

Recently, the ATLAS~\cite{ATLAS:2025evx} and CMS~\cite{CMS:2025ttk} analyses have significantly improved the sensitivity to slepton coannihilation scenarios.
Both collaborations analyzed proton–proton collision data at $\sqrt{s}=13$~TeV, 
corresponding to integrated luminosities of 138~fb$^{-1}$ (CMS) and 140~fb$^{-1}$ (ATLAS). 
They performed dedicated soft-lepton searches in association with a hard initial-state radiation (ISR) jet, optimized for compressed mass spectra, targeting the process
\begin{align}
    p\ p \rightarrow \Tilde{\ell}_{L/R}\ \Tilde{\ell}_{L/R} \rightarrow \Tilde{\chi}^0_1\ell\  \Tilde{\chi}^0_1\ell,  
\end{align}
where $\ell = e,\,\mu$. 
The slepton pair system recoils against the ISR jet, producing large missing transverse momentum in the final state.
In this section, we adopt the CMS results~\cite{CMS:2025ttk} for the selectron and smuon channels, since they provide separate limits for left- and right-handed sleptons, whereas the ATLAS analysis~\cite{ATLAS:2025evx} assumes $m_{\tilde{\ell}_L}=m_{\tilde{\ell}_R}$.

The resulting exclusion limits are shown as the orange-shaded regions in 
Figs.~\ref{fig:BR_BL} and~\ref{fig:BR_BL_Universal}. 
These limits are obtained from the CMS soft-lepton search described above, 
and correspond to the 95\%~CL excluded regions.
As can be seen, these searches probe, for the first time, the parameter space consistent with the observed relic abundance in the $\mathcal{O}(100)$~GeV mass range.
Combining these lower bounds from the LHC searches and the upper bounds imposed by the relic abundance requirement discussed in Sec.~\ref{subsec:Bino-Slepton-omega}, 
the viable mass range for the DM and the charged sleptons is tightly constrained to 
$\mathcal{O}(100$–$400)$~GeV, depending on the scenario.
More quantitatively, the allowed mass envelopes are 
\begin{itemize}
    \item BL models:\; 
    $m_{\tilde{\chi}_1^0}\simeq 170$ -- $420$~GeV,\;
    $m_{\tilde{\ell}_L} \simeq 190$ -- $430$~GeV,
    \item BR models:\; 
    $m_{\tilde{\chi}_1^0}\simeq 130$ -- $430$~GeV,\;
    $m_{\tilde{\ell}_R} \simeq 140$ -- $430$~GeV,
\end{itemize}
where the weakest LHC lower bounds are obtained for $\tilde{\ell}=\tilde{\mu}$, whereas the upper ends of the envelopes set by the relic-abundance requirement are scenario-dependent.
In particular, the weakest upper bound is found for $\tilde{\ell}=\tilde{e}$ and $\tilde{\mu}$ and $\tan\beta=40$ in the BL case, 
while it is found for the universal $\tilde{\ell}_R$ case with $\tan\beta=3$ in the BR case.

\section{Slepton Coannihilation Model with Light Higgsino}
\label{sec:Bino-Slepton-Higgsino}

In Section \ref{sec:Bino-Slepton}, we discussed the current constraints on the minimal slepton coannihilation model, in which all SUSY particles except the bino and sleptons are decoupled.
In this section, we relax this assumption and consider a scenario where the higgsino is also at the electroweak scale.
This setup is well motivated from the viewpoint of the naturalness of the electroweak symmetry breaking~\cite{Barbieri:1987fn,Kitano:2006gv,Baer:2012up}.

In Sec.~\ref{subsec:Bino-Slepton-Higgsino}, we describe the model setup and discuss the constraints from the relic abundance, LHC searches, and dark matter direct detection experiments.
In Sec.~\ref{subsec:g-2} we study the implications of this scenario for the muon $g-2$.

\subsection{Model Setup and Constraints}
\label{subsec:Bino-Slepton-Higgsino}

We consider a slepton coannihilation scenario similar to that discussed in Sec.~\ref{sec:Bino-Slepton}, 
but now allow the higgsino to be light.
Apart from this modification, the model setup and assumptions follow those of the minimal scenario.
As in Sec.~\ref{sec:Bino-Slepton}, the lightest neutralino is predominantly bino-like and plays the role of dark matter.\footnote{We assume $|\mu|>M_1$ and do not consider the parameter region where the lightest neutralino becomes higgsino-like.}
Depending on whether the left-handed or right-handed sleptons are light, we refer to the corresponding scenarios as
the BHL and BHR models, respectively.

The relevant parameters are the same as those in Eq.~\eqref{eq:MSSMparameters}, i.e.,
$M_1$, $M_{L_i/R_i}$, $\tan\beta$, and $\mu$.
For simplicity, in this section, we restrict ourselves to the flavor-universal slepton masses,
$M_{L_1}=M_{L_2}=M_{L_3}\equiv M_L$ and $M_{R_1}=M_{R_2}=M_{R_3}\equiv M_R$.
We take $M_1>0$ and consider both signs of $\mu$.
All the other SUSY particles and the heavy Higgs bosons are decoupled by setting their soft masses to 5~TeV, and we neglect the trilinear $A$-terms and the complex phases of the soft parameters.

\begin{figure}[ht]
  \centering
  \includegraphics[height=5.2cm,angle=0]{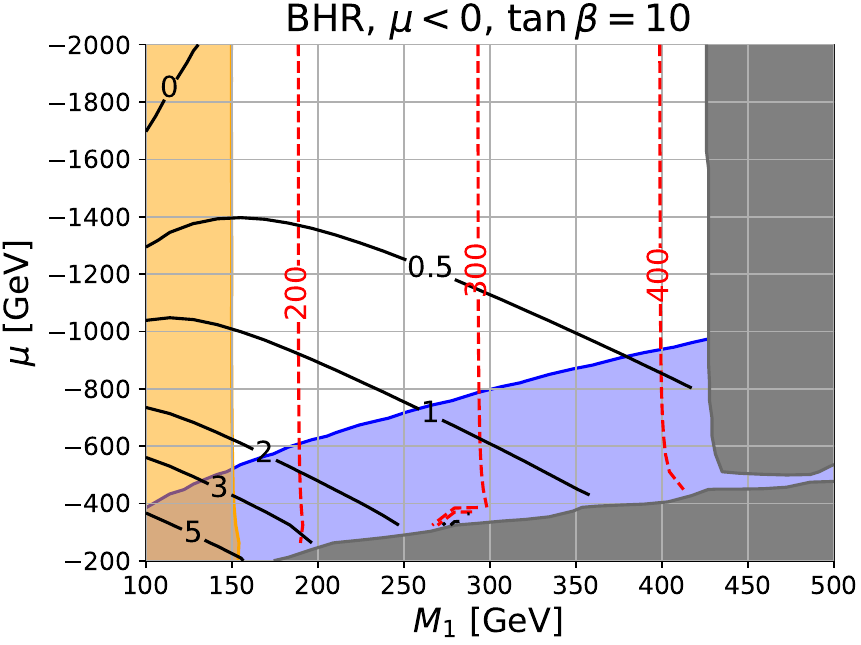}
  \includegraphics[height=5.2cm,angle=0]{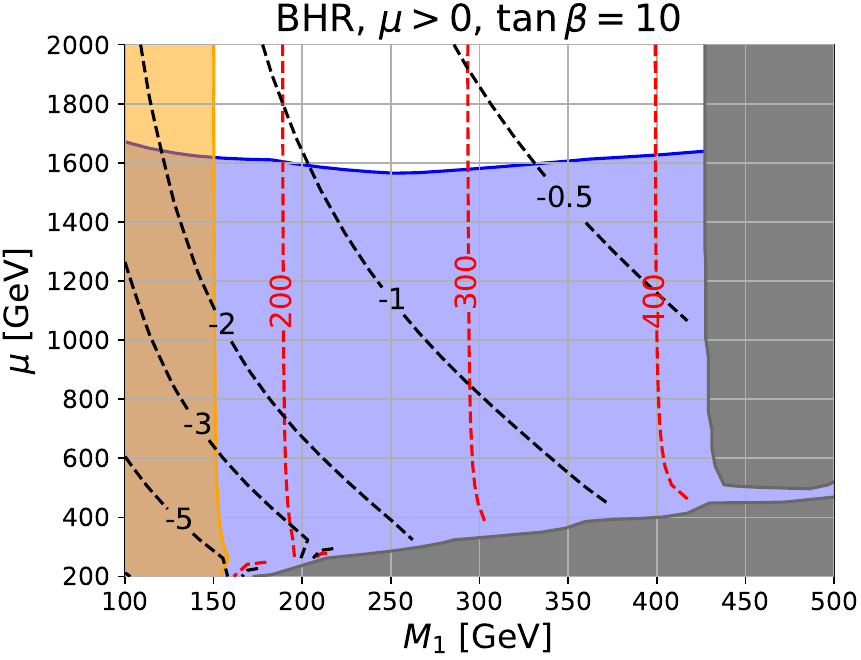}
  \\  
  \includegraphics[height=5.2cm,angle=0]{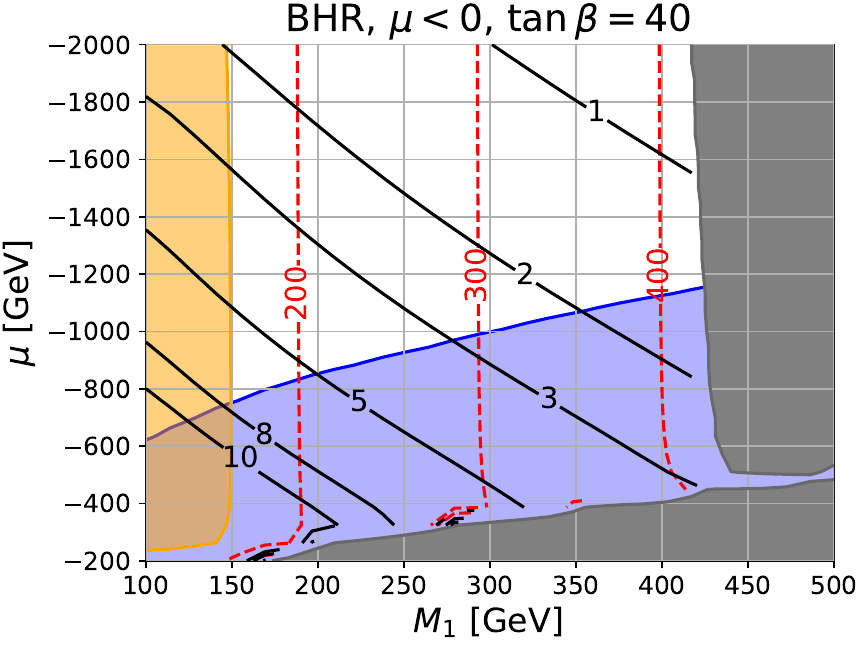}  
  \includegraphics[height=5.2cm,angle=0]{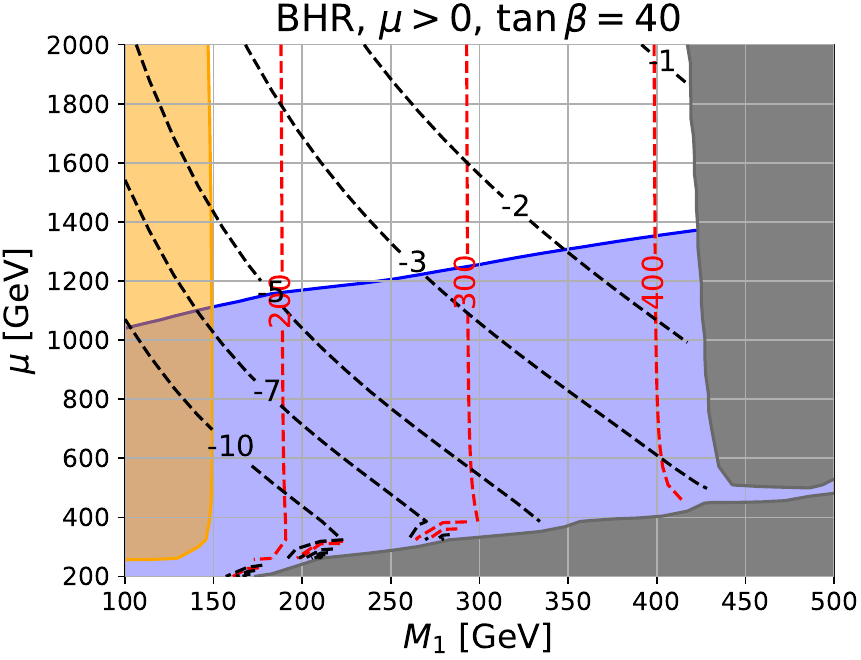}
  \caption{
  Experimental constraints for the BHR model in the $(M_1,|\mu|)$ plane with $\tan\beta=10,40$ (upper and lower panels) and both signs of $\mu$ 
  (left panels: $\mu<0$, right panels: $\mu>0$).
  At each point in the plane, the right-handed slepton soft mass $M_R$ is chosen to reproduce the observed relic abundance $\Omega h^{2}=0.120$ with a bino-like neutralino 
  as the DM, and the red dashed contours indicate the corresponding physical right-handed slepton masses $m_{\tilde{\ell}_R}$ ($\ell=e,\mu$) in GeV.
  The navy shaded region is excluded by the LZ experiment ($90\%$ CL)~\cite{LZ:2024zvo}, while the orange shaded region is excluded by the CMS slepton search~\cite{CMS:2025ttk}.
  The greyed-out region corresponds to the parameter space where no solution reproducing the observed relic abundance exists. 
  The solid (dashed) black contours represent positive (negative) values of the SUSY contribution to the muon $g-2$, $\Delta a_\mu^{\rm SUSY}$, in units of $10^{-10}$.}
  \label{fig:BHR}
\end{figure}

\begin{figure}[t]
  \centering
  \includegraphics[height=5.2cm,angle=0]{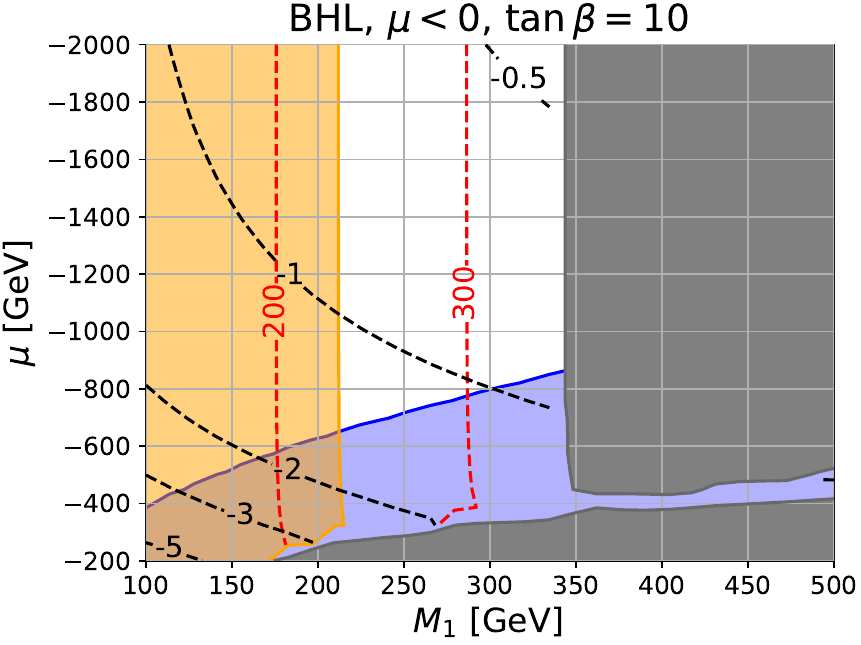}
  \includegraphics[height=5.2cm,angle=0]{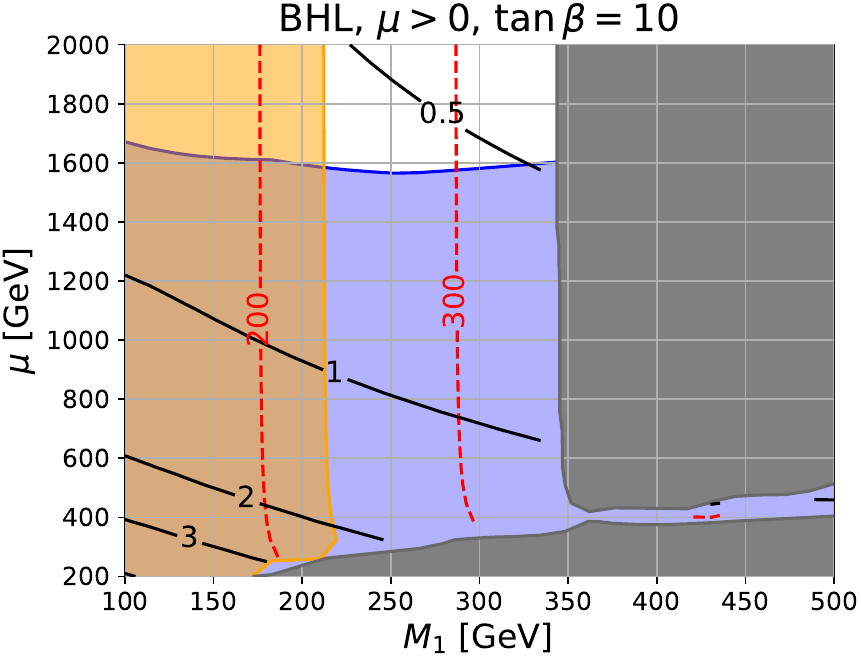}
  \\  
  \includegraphics[height=5.2cm,angle=0]{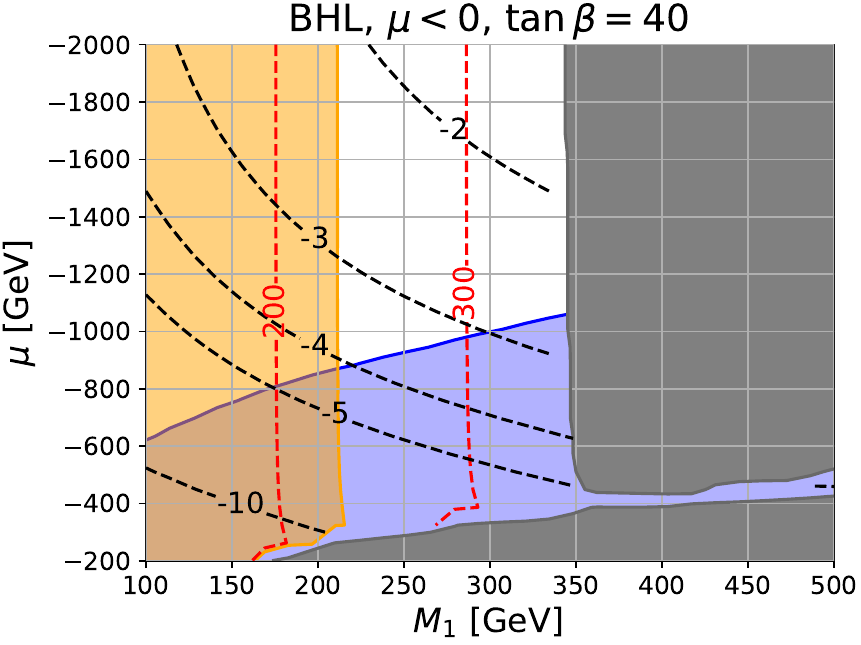}  
  \includegraphics[height=5.2cm,angle=0]{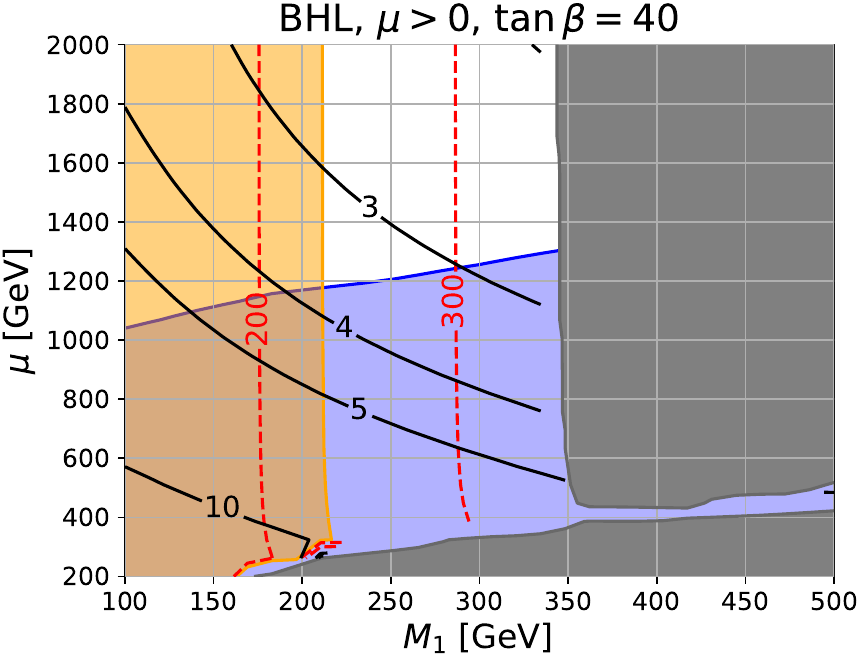}
  \caption{
  Same as Fig.~\ref{fig:BHR}, but for the BHL model with light left-handed sleptons. 
  The red dashed contours indicate the values of $m_{\tilde{\ell}_L}$ ($\ell=e,\mu$) required to reproduce the observed relic abundance, while all other conventions are the same as in Fig.~\ref{fig:BHR}.}
  \label{fig:BHL}
\end{figure}

\begin{figure}[ht]
    \centering
    \includegraphics[height=5.2cm,angle=0]{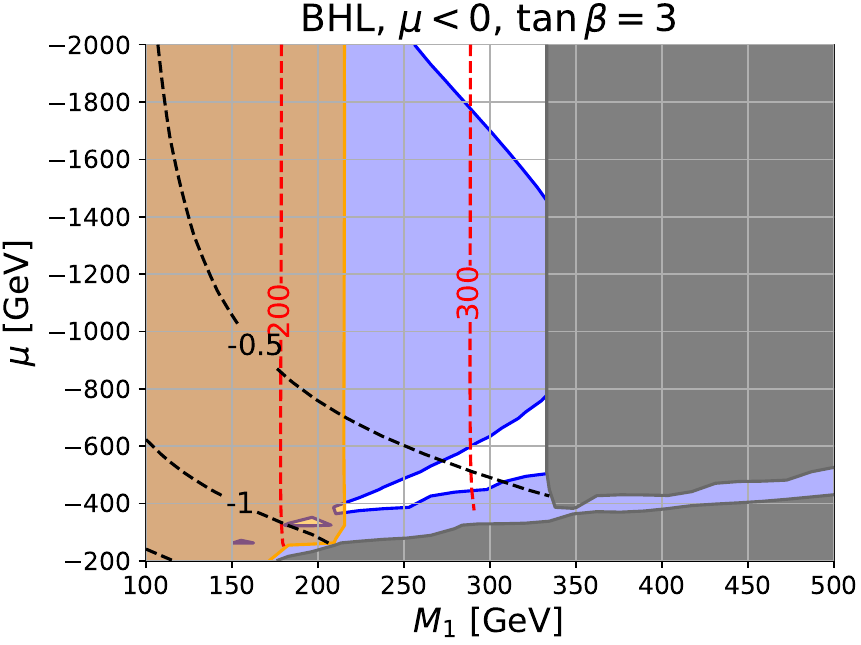}
    \includegraphics[height=5.2cm,angle=0]{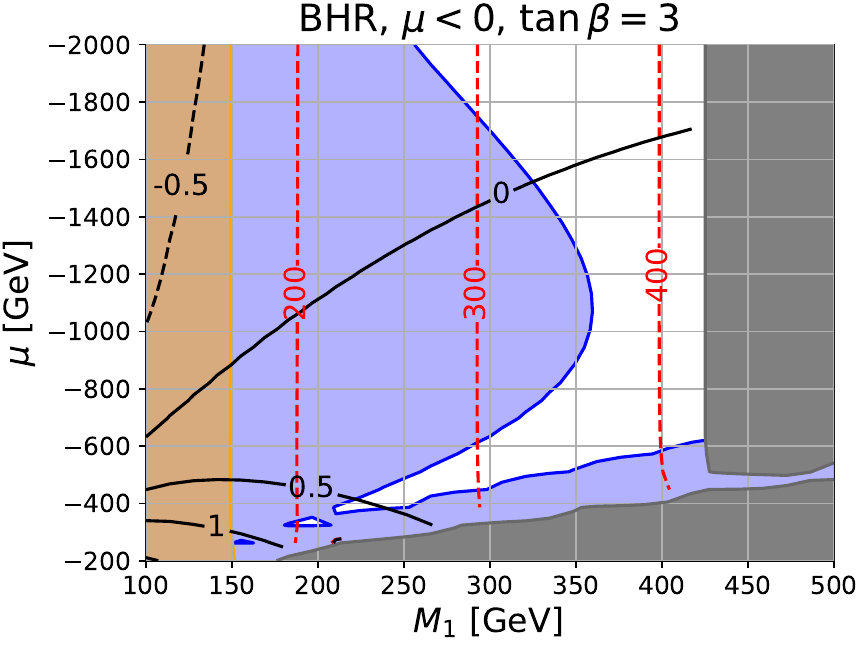}
    \caption{
    Same as Figs.~\ref{fig:BHR} and \ref{fig:BHL}, but for $\tan\beta = 3$ and $\mu<0$.
    The left (right) panel corresponds to the BHL (BHR) model. 
    }
    \label{BHLRtanB=3}
\end{figure}

Our results are summarized in Figs.~\ref{fig:BHR}, \ref{fig:BHL}, and \ref{BHLRtanB=3}.
These figures show the constraints from the relic abundance, LHC searches, and dark matter direct detection experiments in the $(M_1,|\mu|)$ plane, which are discussed below.
We also display contours of the SUSY contribution to the muon $g-2$, which will be discussed in Sec.~\ref{subsec:g-2}.
Figs.~\ref{fig:BHR} and \ref{fig:BHL} show the results for the BHR and BHL models for $\tan\beta=10,40$ and both signs of $\mu$, while Fig.~\ref{BHLRtanB=3} presents the corresponding results for $\tan\beta=3$ with $\mu<0$.
The rationale for isolating the $\tan\beta=3$ case is provided below.

At each point in the figures, the slepton mass is fixed so as to reproduce the observed relic abundance,
$\Omega h^2 = 0.120$, with a bino-like neutralino LSP.
The red dashed contours indicate the corresponding 
physical slepton masses $m_{\tilde{\ell}_{L/R}}$ ($\ell=e,\mu$)
to achieve this relic abundance.
These contours are almost vertical and show only a very weak dependence on the sign and magnitude of $\mu$ or on $\tan\beta$, reflecting the fact that the relic abundance is primarily controlled by the mass difference between the bino and the slepton. 
The greyed-out regions correspond to parameter points where the observed relic abundance cannot be reproduced.
At large $M_1$, the annihilation and coannihilation rates are insufficient even for vanishing mass splittings,
leading to an overabundant relic density, as discussed in Sec.~\ref{subsec:Bino-Slepton-omega}.
On the other hand, for small $|\mu|$, the higgsino admixture becomes too large and the relic abundance falls below the observed value.\footnote{Near the boundary between the overabundant and underabundant regions
(at large $M_1$ and small $|\mu|$), there exist narrow parameter regions where these two effects cancel
and the observed relic abundance can be reproduced. These regions are, however, excluded by the LZ direct detection bound.}

The orange-shaded regions in Figs.~\ref{fig:BHR}, \ref{fig:BHL}, and \ref{BHLRtanB=3} are excluded by the CMS soft-lepton search~\cite{CMS:2025ttk}, as described in Sec.~\ref{subsec:Bino-Slepton-exp}.
These exclusion regions are also almost vertical and show only a very weak dependence on the sign and magnitude of $\mu$ or on $\tan\beta$.
This is because the collider sensitivity is determined primarily by the mass spectrum of the lightest neutralino and the charged slepton, and is essentially insensitive to the higgsino sector.

The navy-shaded regions in Figs.~\ref{fig:BHR}, \ref{fig:BHL}, and \ref{BHLRtanB=3}
are excluded by the LZ direct detection experiment~\cite{LZ:2024zvo}.
We compute the spin-independent dark matter-nucleon scattering cross section using {\tt micrOMEGAs 6.2.3}~\cite{Belanger:2001fz,Alguero:2023zol} and compare it with the LZ limit~\cite{LZ:2024zvo}.

In the BHL and BHR scenarios, direct detection experiments become relevant due to the presence of a sizable higgsino component in the dark matter, which induces spin-independent scattering via Higgs exchange.\footnote{The bounds from LZ can exhibit a mild dependence on the mass of the heavy CP-even Higgs. The lightest neutralino has interactions with quarks via the Standard Model Higgs and the heavy CP-even Higgs. For negative $\mu$, the interference of the amplitudes of these two exchanges is destructive, and so the constraints from the direct detection are weaker \cite{Huang:2014xua}. The current lower limit of heavy Higgs mass is $\sim1800$ GeV \cite{ATLAS:2020zms}, and we have found that CP-even heavy Higgs in this scale can weaken the bound on the chargino mass about 100 GeV, especially for large $\tan\beta$ cases.}
The interaction between the dark matter and Higgs is given by the following Lagrangian:
\begin{equation}
    \mathcal{L} \ni \frac{1}{2}\lambda_{h}h\bar{\tilde{\chi}}_{1}^{0}\tilde{\chi}_{1}^{0}
\end{equation}
where $h$ denotes the Standard Model Higgs boson, and the Higgs–neutralino coupling $\lambda_h$ is approximated as~(cf.~\cite{Hamaguchi:2015rxa})
\begin{equation}
\label{Nh}
    \lambda_{h} \approx g_{1}\left(
    \frac{\mu\sin(2\beta)+M_{1}}{\mu^{2}-M_{1}^{2}}\, m_{Z} \sin\theta_W
    + \mathcal{O}\!\left(\frac{m_{Z}\sin\theta_{W}}{\mu}\right)^3
    \right),
\end{equation}
where $g_1$ and $\theta_W$ are the $U(1)_Y$ gauge coupling and the weak mixing angle, respectively.
The leading term depends on the combination $\mu\sin(2\beta)+M_1$, and therefore exhibits a non-trivial dependence on the sign of $\mu$ as well as on $\tan\beta$.
As a result, the direct detection constraints are generically weaker for $\mu<0$ than for $\mu>0$, as seen in the figures.
Since this coupling does not involve sleptons, the LZ bound is largely insensitive to whether the light slepton is left- or right-handed, and depends mainly on $\mu$ and $\tan\beta$.

From Figs.~\ref{fig:BHR} and \ref{fig:BHL}, we observe that the LZ constraints for positive $\mu$, represented by the navy regions, become more stringent as $\tan\beta$ decreases. 
This is consistent with the fact that the leading term in Eq.~\eqref{Nh} increases as $\tan\beta$ decreases through $\sin(2\beta)$. Indeed, when plotting the LZ constraints for $\tan\beta = 3$ and $\mu>0$, one finds that the entire parameter space for $\mu<2$~TeV would be excluded. Therefore, we omit the plots for $\tan\beta = 3$ when $\mu$ is positive.

The case for $\tan\beta = 3$ and $\mu < 0$ is shown in Fig.~\ref{BHLRtanB=3}.
In this case, the leading Higgs-neutralino coupling in Eq.~\eqref{Nh} vanishes at
$\mu = -M_1/\sin(2\beta)$,
giving rise to a dark matter direct detection blind spot~\cite{Cheung:2012qy}.
For $\tan\beta = 3$, this corresponds to $\mu = -5M_1/3$,
which lies in the parameter region shown in Fig.~\ref{BHLRtanB=3}.
Away from this blind-spot region, the spin-independent dark matter scattering cross section increases, leaving only a narrow blind-spot band that is not excluded by the LZ constraints.
For even larger values of $|\mu|$, the Higgs–neutralino coupling is again suppressed, leading to a weakening of the direct detection constraint.

In summary, when the relic abundance, LHC, and direct detection constraints are combined,
we find they exclude complementary regions of the parameter space.
The CMS soft-lepton search, shown by the orange-shaded regions, excludes the low-$M_1$ region,
while the relic abundance requirement removes the large-$M_1$ region.
The LZ direct detection bounds, depicted by the navy-shaded regions, primarily exclude the small-$|\mu|$ region,
with additional structure appearing for $\tan\beta = 3$ due to the direct detection blind spot.
The remaining white regions in the figures are not constrained by current experiments
and constitute viable parameter space that remains an important target for future collider
and dark matter search experiments.

As a final remark in this subsection, we briefly comment on additional constraints beyond those shown in the figures. 
First, we find that the present-day annihilation cross section of DM is typically ${\cal O}(10^{-28})\,\mathrm{cm}^3/\mathrm{s}$ or smaller in the viable parameter regions shown in Figs.~\ref{fig:BHR}--\ref{BHLRtanB=3}, although it can reach ${\cal O}(10^{-27})\,\mathrm{cm}^3/\mathrm{s}$ in a small region near the blind spot. In all cases, it remains well below current indirect-detection bounds~\cite{Cirelli:2024ssz}.
Next, we discuss possible LHC constraints from higgsino pair production.
At the LHC, depending on the higgsino decay modes, the following channels are available:
\begin{align}
    pp &\rightarrow \Tilde{\chi}^0_{2/3}\ \Tilde{\chi}^\pm \rightarrow (\Tilde{\chi}^0_1Z)(\Tilde{\chi}^0_1W), \nonumber\\
    pp &\rightarrow \Tilde{\chi}^0_{2/3}\ \Tilde{\chi}^\pm \rightarrow (\Tilde{\chi}^0_1h)(\Tilde{\chi}^0_1W), \nonumber\\
    pp &\rightarrow \Tilde{\chi}^0_{2/3}\ \Tilde{\chi}^\pm \rightarrow (\Tilde{\tau}\tau)(\Tilde{\nu}\tau) \rightarrow(\Tilde{\chi}^0_1\tau\tau)(\Tilde{\chi}^0_1\nu\tau), \nonumber\\
    pp &\rightarrow \Tilde{\chi}^\pm\ \Tilde{\chi}^\pm  \rightarrow (\Tilde{\nu}\tau)(\Tilde{\nu}\tau) \rightarrow(\Tilde{\chi}^0_1\nu\tau)(\Tilde{\chi}^0_1\nu\tau),\nonumber\\
    pp &\rightarrow \Tilde{\chi}^0_{2}\ \Tilde{\chi}^0_3 \rightarrow (\Tilde{\tau}\tau)(\Tilde{\tau}\tau) \rightarrow(\Tilde{\chi}^0_1\tau\tau)(\Tilde{\chi}^0_1\tau\tau).
    \label{eq:LHCtautau}
\end{align}
The branching ratios depend on $\tan\beta$ through the tau Yukawa coupling, which enhances decays into taus at large $\tan\beta$.
We have recast the LHC results \cite{ATLAS:2019gti,ATLAS:2017qwn,ATLAS:2020pgy,ATLAS:2021moa,ATLAS:2021yqv,ATLAS:2022zwa} into the parameter space of the BHL and BHR models, following the previous analyses \cite{Endo:2020mqz,Endo:2017zrj}. 
The most stringent limits come from the $2\tau$ channels, namely the last three channels in Eq.\eqref{eq:LHCtautau}, especially for large $\tan\beta$~\cite{Endo:2017zrj}.
However, these constraints are still weaker than those obtained by combining the CMS soft lepton analysis and the LZ experiment.
Therefore, we have not included these constraints in the figures.

\subsection{Implication for the muon $g-2$}
\label{subsec:g-2}

The muon anomalous magnetic moment $a_\mu \equiv (g-2)_\mu/2$ is one of the most sensitive probes of new physics.
The latest result from the Fermilab Muon $g-2$ Collaboration leads to the world average~\cite{Muong-2:2025xyk}
\begin{align}
    a_\mu({\rm exp}) = (11,659,207.15 \pm 1.45) \times 10^{-10}.
\end{align}
The updated SM prediction, based on lattice-QCD determinations of the leading-order hadronic vacuum polarization contribution, is~\cite{Aliberti:2025beg}
\begin{align}
    a_\mu({\rm theory}) = (11,659,203.3 \pm 6.2) \times 10^{-10},
\end{align}
which corresponds to
\begin{align}
    \Delta a_\mu \equiv a_\mu({\rm exp})-a_\mu({\rm theory})
    = (3.8 \pm 6.3)\times 10^{-10}.
    \label{eq:Delta_a_mu_latest}
\end{align}
This indicates that the experimental result is consistent with the SM within current uncertainties.

In this work, we present the SUSY contribution $\Delta a_\mu^{\rm SUSY}$ in the parameter space of interest. 
Since the sign of $\Delta a_\mu^{\rm SUSY}$ depends on the sign of $\mu$ in our setup, we consider both signs of $\mu$ and show the corresponding $\Delta a_\mu$ contours.
For a recent review on the muon $(g-2)_\mu$ as a probe of physics beyond the Standard Model, see Ref.~\cite{Athron:2025ets}.
See also Refs.~\cite{Endo:2017zrj,Endo:2021zal} for related MSSM studies of the interplay among dark matter phenomenology, the SUSY contribution to the muon $g-2$, and LHC searches.

In Figs.~\ref{fig:BHR},
\ref{fig:BHL}, and \ref{BHLRtanB=3},
the SUSY contributions to the muon $g-2$ in the BHL and BHR scenarios are shown by solid and dashed black contours 
for $\Delta a_\mu^{\rm SUSY}>0$ and $\Delta a_\mu^{\rm SUSY}<0$, respectively.
For the numerical evaluation of $\Delta a_\mu^{\rm SUSY}$, we use the public code~{\tt GM2Calc} \cite{Athron:2015rva,Athron:2021evk},
which computes the MSSM contribution at full one-loop level.\footnote{We do not include the two-loop corrections, since most superparticles are decoupled in our setup and a consistent inclusion of higher-order effects would go beyond our simplified framework.}

In the mass-insertion approximation applicable to large values of $\tan\beta$, the leading contributions to $\Delta a_\mu$ in the BHL and BHR scenarios are given by~\cite{Moroi:1995yh}: 
\begin{align}
\label{g-2BHL}
\Delta a_\mu^{\textrm{BHL}} &\simeq 
\frac{\alpha_Y}{8\pi}\frac{m_\mu^2}{M_1\mu}\tan{\beta}\:f_N\left(\frac{M_1^2}{m_{\tilde{\mu},L}^2},\frac{\mu^2}{m_{\tilde{\mu},L}^2}\right), 
\\
\label{g-2BHR}
\Delta a_\mu^{\textrm{BHR}} &\simeq
-\frac{\alpha_Y}{4\pi}\frac{m_\mu^2}
{M_1\mu}\tan{\beta}\: f_N\left(\frac{M_1^2}{m_{\tilde{\mu},R}^2},\frac{\mu^2}{m_{\tilde{\mu},R}^2}\right),
\end{align}
where the loop-function $f_{N}$ is given by
\begin{align}
    f_N (x,y) &= xy\left[\frac{-3+x+y+xy}{(x-1)^2(y-1)^2} + \frac{2x\ln x}{(x-y)(x-1)^3} - \frac{2y\ln y}{(x-y)(y-1)^3}\right],
\end{align}
and satisfies $0\le f_N(x,y)\le 1$.
Therefore, the sign of $\Delta a_\mu^{\rm SUSY}$ is set by the sign of $\mu$: it is positive (negative) for $\mu>0$ in the BHL (BHR) scenario, and flips for $\mu<0$. The magnitude scales approximately as $|\Delta a_\mu^{\rm SUSY}|\propto \tan\beta/(|M_1\mu|)$ (up to the loop function), and is thus enhanced at large $\tan\beta$ and suppressed for large $|\mu|$ and/or $M_1$.

These qualitative behaviors are reflected in the contours shown in the figures with $\tan\beta = 10,\: 40$, i.e., Figs.~\ref{fig:BHR} and \ref{fig:BHL}.
We find that the parameter regions yielding large $|\Delta a_\mu^{\rm SUSY}|$ are largely excluded once the relic-abundance requirement, the LHC soft-lepton search, and the LZ bound are imposed; within the parameter space that survives all constraints in Sec.~\ref{subsec:Bino-Slepton-Higgsino}, $\Delta a_\mu^{\rm SUSY}$
is limited as
\begin{align}
& 
-4\times 10^{-10} \le \Delta a_\mu^{\textrm{BHL}} \le 4\times 10^{-10},
\label{eq:a_mu_range_BHL}
\\
&-7\times 10^{-10} \le \Delta a_\mu^{\textrm{BHR}} \le 8 \times 10^{-10}.
\label{eq:a_mu_range_BHR}
\end{align}
These ranges are compatible with Eq.~\eqref{eq:Delta_a_mu_latest}
within $2\sigma$.

So far, we have focused on the BHL and BHR scenarios discussed in Sec.~\ref{subsec:Bino-Slepton-Higgsino}.
In addition to these scenarios, we now briefly discuss the BLR scenario for completeness,
in which both the left- and right-handed sleptons are light and degenerate in mass.
This scenario can also accommodate bino-slepton coannihilation dark matter
and may lead to sizable contributions to $\Delta a_\mu$~\cite{Endo:2013lva,Endo:2021zal}.

\begin{figure}[t]
    \centering
    \includegraphics[height=5.2cm,angle=0]{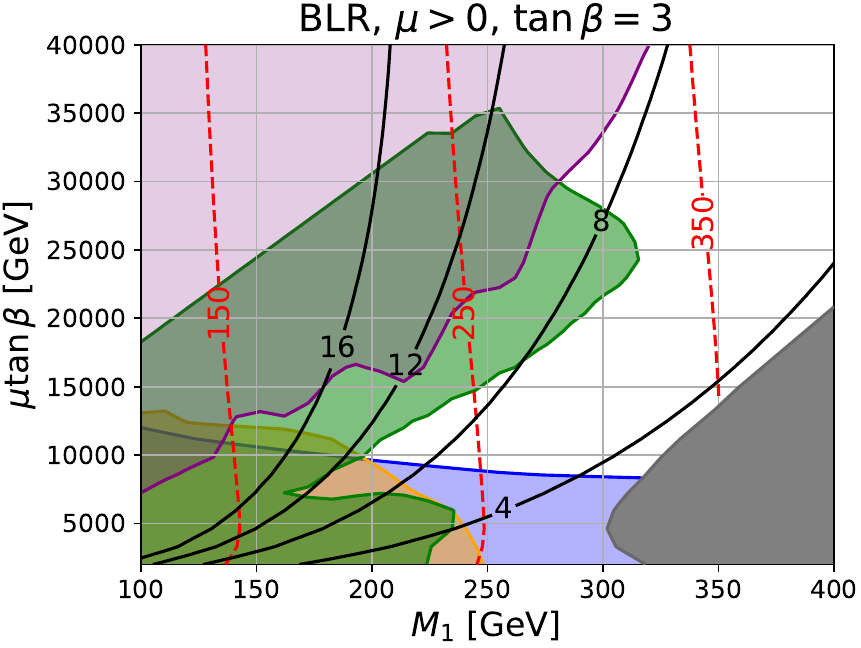}
    \includegraphics[height=5.2cm,angle=0]{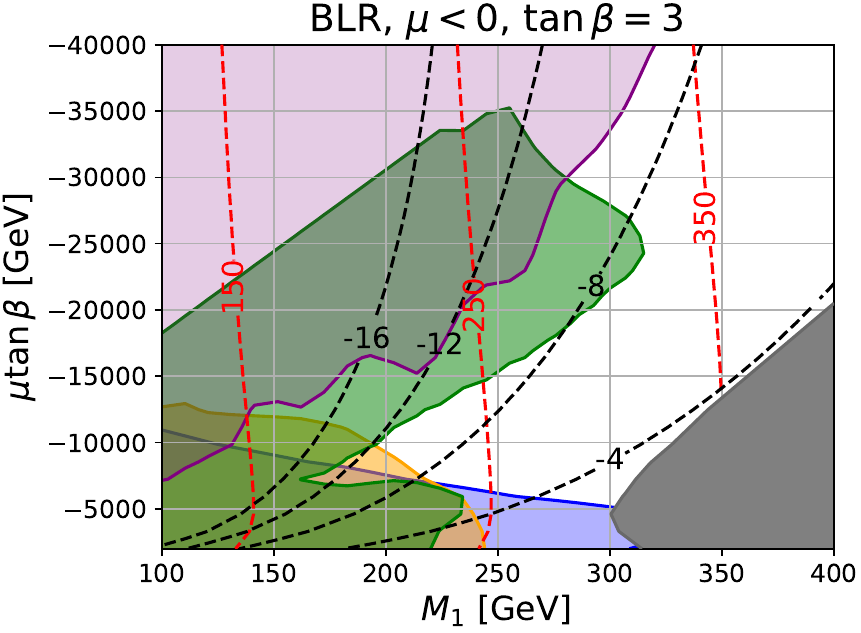}   
    \caption{Constraints and relevant contours for the BLR model with $\tan{\beta}=3$ in the $(M_1, |\mu|\tan\beta)$-plane. The red dotted lines represent the lighter stau  masses in GeV that give $\Omega h^{2} = 0.120$. 
    The solid (dashed) black contours represent positive (negative) values of the SUSY contribution to the muon $g-2$, $\Delta a_\mu^{\rm SUSY}$, in units of $10^{-10}$.
    The coloured regions represent parameter spaces excluded by the recent soft-lepton searches (orange: CMS~\cite{CMS:2025ttk}, green: ATLAS~\cite{ATLAS:2025evx}), the conventional slepton search (purple: ATLAS~\cite{ATLAS:2019lff}), and the LZ results (navy~\cite{LZ:2024zvo}).
    The greyed-out region refers to the parameter space with no solution that could reproduce a relic abundance of $\Omega h^{2} = 0.120$ with the bino-like neutralino as the LSP.
    }
    \label{fig:BLR}
\end{figure}

In Fig.~\ref{fig:BLR}, we show the contours of the SUSY contribution to the muon $g-2$ in the $(M_1, |\mu|\tan\beta)$-plane for $\tan\beta=3$.
As in Sec.~\ref{subsec:Bino-Slepton-Higgsino}, at each point in the figures, the slepton masses are fixed so as to reproduce the observed relic abundance,
$\Omega h^2 = 0.120$, with a bino-like neutralino LSP. Here, we impose $m_{\tilde{\ell}_L}=m_{\tilde{\ell}_R}$ with $\ell=e,\mu$, which allows for a clearer comparison with the LHC constraints and
typically maximizes the SUSY contribution to $\Delta a_\mu$.
The coloured regions indicate the parameter space excluded by the recent soft-lepton searches (orange: CMS~\cite{CMS:2025ttk}, green: ATLAS~\cite{ATLAS:2025evx}), the conventional slepton search (purple: ATLAS~\cite{ATLAS:2019lff}), and the LZ results (navy~\cite{LZ:2024zvo}), while the greyed-out region corresponds to parameter points where no solution reproducing the observed relic abundance exists.
From Fig.~\ref{fig:BLR}, we obtain
\begin{align}
-12\times 10^{-10} \le \Delta a_\mu^{\textrm{BLR}} \le 11 \times 10^{-10}.
\label{eq:amu_range_BLR}
\end{align}
We also find that this range is not enlarged by increasing $|\mu|\tan\beta$ or $\tan\beta$ itself.
The parameter regions with large $|\mu|\tan\beta$ are already excluded by the LHC and vacuum-stability constraints, while varying $\tan\beta$ mainly shifts the LZ bound without enlarging the allowed range of $\Delta a_\mu^{\rm BLR}$.
Details of the BLR scenario are given in Appendix~\ref{app:BLR}.

Comparing Eq.~\eqref{eq:amu_range_BLR} with Eq.~\eqref{eq:Delta_a_mu_latest},
we find that the lower end of the $\Delta a_\mu^{\textrm{BLR}}$ range lies about $2.5\sigma$ away from the current central value.
In particular, the region in Fig.~\ref{fig:BLR} corresponding to $\Delta a_\mu^{\rm BLR}\lesssim -9\times 10^{-10}$ lies outside the $2\sigma$ range of Eq.~\eqref{eq:Delta_a_mu_latest}.

Combining Eqs.~\eqref{eq:a_mu_range_BHL},  \eqref{eq:a_mu_range_BHR}, and \eqref{eq:amu_range_BLR},  we conclude that the SUSY contribution to the muon $g-2$ is limited to 
\begin{align}
-12\times 10^{-10} \le \Delta  a_\mu^{\rm SUSY} \le 11 \times 10^{-10}
\end{align}
in the BHL, BHR, and BLR scenarios with flavor-universal slepton soft masses realizing bino-slepton coannihilation dark matter,  
once the relic-abundance requirement, the LHC constraints, and the dark matter direct detection bound are imposed.

\section{Summary and Discussion}
\label{sec:summary}

We have revisited the bino-slepton coannihilation dark matter scenario in supersymmetric models in light of recent experimental developments.
We first studied the minimal setup in which only a bino-like neutralino and either left- or right-handed sleptons are light, while all other superparticles are decoupled.
Using the latest LHC searches for compressed spectra, we have shown that the viable parameter space of this scenario is significantly constrained, restricting the dark matter mass to about 170--420 GeV for left-handed slepton coannihilation and 130--430 GeV for right-handed slepton coannihilation.

We then extended the analysis to scenarios with a light higgsino.
In this case, direct detection experiments, particularly the LZ results, impose strong constraints on the higgsino mass.
We have also examined the implications for the muon anomalous magnetic moment in the BHL, BHR, and BLR scenarios and found that, once the relic-abundance requirement and current collider and direct-detection limits are imposed, the supersymmetric contribution is limited as $|\Delta a_\mu^{\rm SUSY}| \lesssim 1.2\times10^{-9}$.

Bino-slepton coannihilation remains a viable and well-motivated thermal relic dark matter scenario.
The mass range identified in this paper provides a clear target for future collider searches.
Future lepton colliders represent one of the most promising approaches, as their clean environment is particularly advantageous for probing nearly degenerate spectra.
For example, linear colliders such as the ILC~\cite{ILCInternationalDevelopmentTeam:2022izu} with a center-of-mass energy of 500 GeV are expected to probe slepton masses up to about 250 GeV even when the mass splitting is below $\sim 10$ GeV.
In addition, future high-luminosity circular colliders may also probe this scenario indirectly through precision electroweak measurements.
A detailed study of such probes, as well as their implications for more general coannihilation dark matter scenarios, is left for future work.

\section*{Acknowledgments}

We thank Motoi Endo, Sho Iwamoto, and Teppei Kitahara for collaboration at an early stage of this work and for helpful comments on the manuscript.
We also thank Shion Chen for useful discussions.
This work was supported by JSPS KAKENHI Grant Numbers 24H02244 and No. 24K07041.

\appendix
\section{BLR model}
\label{app:BLR}

In Sec.~\ref{sec:Bino-Slepton-Higgsino}, we mainly discussed the BHL and BHR scenarios.
In this appendix, we consider the BLR scenario, in which both left- and right-handed sleptons are light.
As in Sec.~\ref{sec:Bino-Slepton-Higgsino}, we assume that a bino-like neutralino LSP constitutes the dark matter.
The relevant parameters are the same as those in Eq.~\eqref{eq:MSSMparameters}, i.e.,
$M_1$, $M_{L_i/R_i}$, $\tan\beta$, and $\mu$.
For simplicity, we restrict ourselves to the flavor-universal slepton soft masses,
$M_{L_1}=M_{L_2}=M_{L_3}\equiv M_L$ and $M_{R_1}=M_{R_2}=M_{R_3}\equiv M_R$.
We also impose that the physical masses of the left- and right-handed selectrons and smuons are equal, $m_{\tilde{e}_L}=m_{\tilde{e}_R}=m_{\tilde{\mu}_L}=m_{\tilde{\mu}_R}$, which allows for a direct comparison with the LHC constraints and typically maximizes the SUSY contribution to $\Delta a_\mu$.
As in Sec.~\ref{sec:Bino-Slepton-Higgsino}, we decouple the remaining SUSY particles and the heavy Higgs bosons by setting their soft masses to $5$~TeV, while neglecting the trilinear $A$-terms and possible complex phases of the soft parameters.

The $g-2$ contributions from BLR scenarios in large $\tan\beta$ can be approximated by~\cite{Moroi:1995yh}:
\begin{equation}
\label{g-2BLR}
    \Delta a_{\mu}^{\text{BLR}}\simeq \frac{\alpha_{Y}}{4\pi}\frac{m_{\mu}^{2}M_{1}}{m_{\tilde{\mu}_{L}}^{2}m_{\tilde{\mu}_{R}}^{2}}\mu\tan\beta\cdot f_{N}\Bigg(\frac{m_{\tilde{\mu}_{L}}^{2}}{M_{1}^{2}},\frac{m_{\tilde{\mu}_{R}}^{2}}{M_{1}^{2}}\Bigg)
\end{equation}
In contrast to Eqs.~\eqref{g-2BHL} and \eqref{g-2BHR} for the BHL and BHR scenarios,
the contribution in Eq.~\eqref{g-2BLR} depends on the combination $\mu\tan\beta$ rather than on $\tan\beta/\mu$.
Therefore, large values of $|\mu|\tan\beta$ can lead to sizable contributions to $\Delta a_\mu^{\rm SUSY}$.

In Fig.~\ref{fig:BLR}, we show the contours of the SUSY contribution to the muon $g-2$
in the $(M_1, |\mu|\tan\beta)$ plane for $\tan\beta=3$.
At each point in the figure, the relevant slepton masses are fixed so as to reproduce the observed relic abundance,
$\Omega h^2 = 0.120$, with a bino-like neutralino LSP.
The red dashed contours represent the physical mass of the lighter stau, which is the dominant coannihilation partner of the bino-like neutralino LSP in the present setup.
The coloured regions indicate the parameter space excluded by the recent soft-lepton searches (orange: CMS~\cite{CMS:2025ttk}, green: ATLAS~\cite{ATLAS:2025evx}), the conventional slepton search (purple: ATLAS~\cite{ATLAS:2019lff}), and the LZ results (navy~\cite{LZ:2024zvo}).

For the soft-lepton search, in addition to the CMS search~\cite{CMS:2025ttk}, the ATLAS search~\cite{ATLAS:2025evx} is applicable because we impose equal physical masses for the left- and right-handed sleptons.
For large $|\mu|\tan\beta$, the conventional slepton search~\cite{ATLAS:2019lff} also becomes relevant for the following reason.
In order to reproduce the observed relic abundance through coannihilation, the lighter stau must remain nearly degenerate with the bino-like neutralino.
Since the left-right mixing of the stau increases with $|\mu|\tan\beta$, this requires larger diagonal entries in the slepton mass matrix [cf.~Eq.~\eqref{eq:slepton_mass_matrix}], which in turn makes the selectron and smuon somewhat heavier than the neutralino.
The corresponding mass splitting is therefore enlarged, and the conventional slepton searches become relevant.

The LZ bound excludes the region with relatively small $|\mu|\tan\beta$.
We have also checked that varying $\tan\beta$ does not enlarge the allowed range of $\Delta a_\mu^{\rm BLR}$, since it mainly shifts the LZ bound while leaving the other relevant contours almost unchanged.

In the BLR scenario, there is one additional important constraint, 
and that is the meta-stability of the electroweak vacuum. 
For large $|\mu|\tan\beta$, the left-right stau mixing becomes sizable, which can induce a deeper charge-breaking minimum.
Focusing on the stau direction, the relevant trilinear coupling to the SM-like Higgs boson is given by~\cite{Endo:2013lva}
\begin{equation}
\label{Vacuum1}
    V\simeq -\frac{m_{\tau}}{\sqrt{2}v(1+\Delta_{\tau})}\mu\tan\beta\cdot\tilde{\tau}_{L}^{*}\tilde{\tau}_{R} h + \text{h.c.}
\end{equation}
where $v\simeq 174$ GeV is the vacuum expectation value of the SM-like Higgs, and $\Delta_{\tau}$ represents the radiative corrections to the tau Yukawa coupling, as discussed in Ref.~\cite{Endo:2013lva}. 
As $|\mu|\tan\beta$ increases, a charge-breaking minimum deeper than the electroweak vacuum can appear.
In that case, the electroweak vacuum becomes metastable and its lifetime may become shorter than the age of the Universe.
Requiring vacuum meta-stability therefore places an upper bound on $|\mu|\tan\beta$.
Using the fitting formula given in Ref.~\cite{Kitahara:2013lfa,Endo:2013lva}, we evaluate this vacuum meta-stability bound on $|\mu|\tan\beta$. 
Although this bound is not shown in Fig.~\ref{fig:BLR}, it becomes relevant for larger values of $|\mu|\tan\beta$ beyond the plotted range.
Combined with the conventional slepton search~\cite{ATLAS:2019lff} discussed above, it prevents $\Delta a_\mu^{\rm BLR}$ from exceeding the range already obtained from Fig.~\ref{fig:BLR}.
Therefore, extending the analysis to larger $|\mu|\tan\beta$ does not enlarge the allowed range of $\Delta a_\mu^{\rm BLR}$ compared with Eq.~\eqref{eq:amu_range_BLR}.


\bibliographystyle{utphysmod}
\bibliography{ref}


\end{document}